\documentclass[a4paper,twocolumn,accepted=2025-05-26]{quantumarticle}
\pdfoutput=1
\usepackage[utf8]{inputenc}
\usepackage[english]{babel}
\usepackage[T1]{fontenc}

\usepackage{graphicx}
\usepackage{dcolumn}
\usepackage{bm}
\usepackage[table]{xcolor}
\usepackage{tabularray}
\usepackage{amsmath}
\usepackage{braket}
\usepackage{subcaption}
\usepackage{hyperref}
\usepackage{upgreek}

\definecolor{tblue}{rgb}{0.121,0.467,0.706}
\definecolor{orange}{rgb}{1,0.498,0.055}
\definecolor{green}{rgb}{0.172,0.627,0.172}
\definecolor{red}{rgb}{0.839,0.153,0.157}
\definecolor{purple}{rgb}{0.58,0.404,0.741}
\definecolor{brown}{rgb}{0.549,0.337,0.294}

\usepackage{xcolor}
\usepackage{soul}
\newcommand{\change}[2]{\iffalse #1 \fi{{#2}}}

\begin{document}

\title{Optimal number of stabilizer measurement rounds in an idling surface code patch}

\author{Áron Márton}%
\affiliation{Department of Theoretical Physics, Institute of Physics, Budapest University of Technology and Economics, Műegyetem rkp. 3., H-1111 Budapest, Hungary}

\author{János K. Asbóth}
\affiliation{Department of Theoretical Physics, Institute of Physics, Budapest University of Technology and Economics, Műegyetem rkp. 3., H-1111 Budapest, Hungary}
\affiliation{HUN-REN Wigner Research Centre for Physics, H-1525 Budapest, P.O. Box 49., Hungary}
\affiliation{HUN-REN-BME-BCE Quantum Technology Research Group, Műegyetem rkp. 3., H-1111 Budapest, Hungary}%


\begin{abstract}
    Logical qubits can be protected against environmental noise by encoding them into a highly entangled state of many physical qubits and actively intervening in the dynamics with stabilizer measurements. In this work, we numerically optimize the rate of these interventions: the number of stabilizer measurement rounds for a logical qubit encoded in a surface code patch and idling for a given time. We model the environmental noise on the circuit level, including gate errors, readout errors, amplitude and phase damping. We find, qualitatively, that the optimal number of stabilizer measurement rounds is getting smaller for better qubits and getting larger for better gates or larger code sizes. We discuss the implications of our results to some of the leading architectures, superconducting qubits, and neutral atoms.
\end{abstract}

\maketitle

\section{Introduction}

Quantum error-correcting codes are necessary for useful quantum algorithms to be run on noisy hardware. One of the most promising candidates among such codes is the surface code \cite{kitaev2003fault,fowler2012surface}, due to its planar connectivity and high threshold \cite{dennis2002topological,Wang_2003,fowler2012surface}. 
Quantum memory experiments with surface codes have been performed on various architectures \cite{google2023suppressing,google_below_threshold,Bluvstein_2022,krinner2022realizing,Marques_2021,Zhao_2022,Abobeih_2022}, 
and logical operations have also been demonstrated recently \cite{Bluvstein_2023,zhang2024demonstrating,hetenyi2024creating}.

The performance of the surface code is routinely investigated under static noise models, i.e., errors occurring with fixed per-cycle probability. This class of errors includes random Pauli and readout errors, which are modeled both on phenomenological \cite{Wang_2003} and circuit level \cite{fowler2012surface}, coherent errors (small unitary rotations) were also investigated both theoretically \cite{Iverson_2020,beale_2018} and numerically \cite{Bravyi_2018,M_rton_2023,pataki2024coherent}.

In a real quantum computer, errors are dynamic: the per-cycle error rate depends on the time between quantum error correction rounds. This raises an optimization problem for a logical qubit based on the surface code: What is the optimal number of stabilizer measurement rounds to get the best performance from a logical qubit over a fixed amount of time? 

In this work, we answer the question of optimizing the number of stabilizer measurement rounds by numerically simulating a circuit-level noise model including static errors (gate and readout errors) and dynamic errors (amplitude and phase damping) as well. We investigate how the optimal number depends on different noise parameters, and on the code distance. We find that, as could be expected, for higher quality physical qubits the optimal number of rounds is lower, while for higher quality gates the optimal number of rounds is higher. Interestingly, the optimal number of rounds is also higher for larger code sizes. We also discuss the practical implications of our results to quantum error correction experiments with superconducting qubits and neutral atoms. Broadly speaking, for superconducting qubits performing as many stabilizer measurement rounds as possible is beneficial; however, for neutral atoms, much fewer measurement rounds\change{}{ may} offer optimal performance.

The rest of this paper is structured as follows. In Sec.~\ref{sec:surface_code} we introduce the basic concepts of the surface code and discuss the logical noise channel corresponding to the logical qubit encoded into a surface code patch. In Sec.~\ref{sec:error} we describe our noise model in detail. Finally, in Sec.~\ref{sec:results} we discuss our results on the optimal number of stabilizer measurement rounds and explain how it depends on the noise parameters and the code distance. We also discuss the implications of our work for superconducting and neutral atom devices.\change{}{ In Sec.~\ref{sec:conclusion} we summarize our results and discuss possible extensions of our work.}

\section{The logical noise channel of the surface code} \label{sec:surface_code}

\begin{figure}[!h]
    \centering
    \includegraphics[width = .45\textwidth]{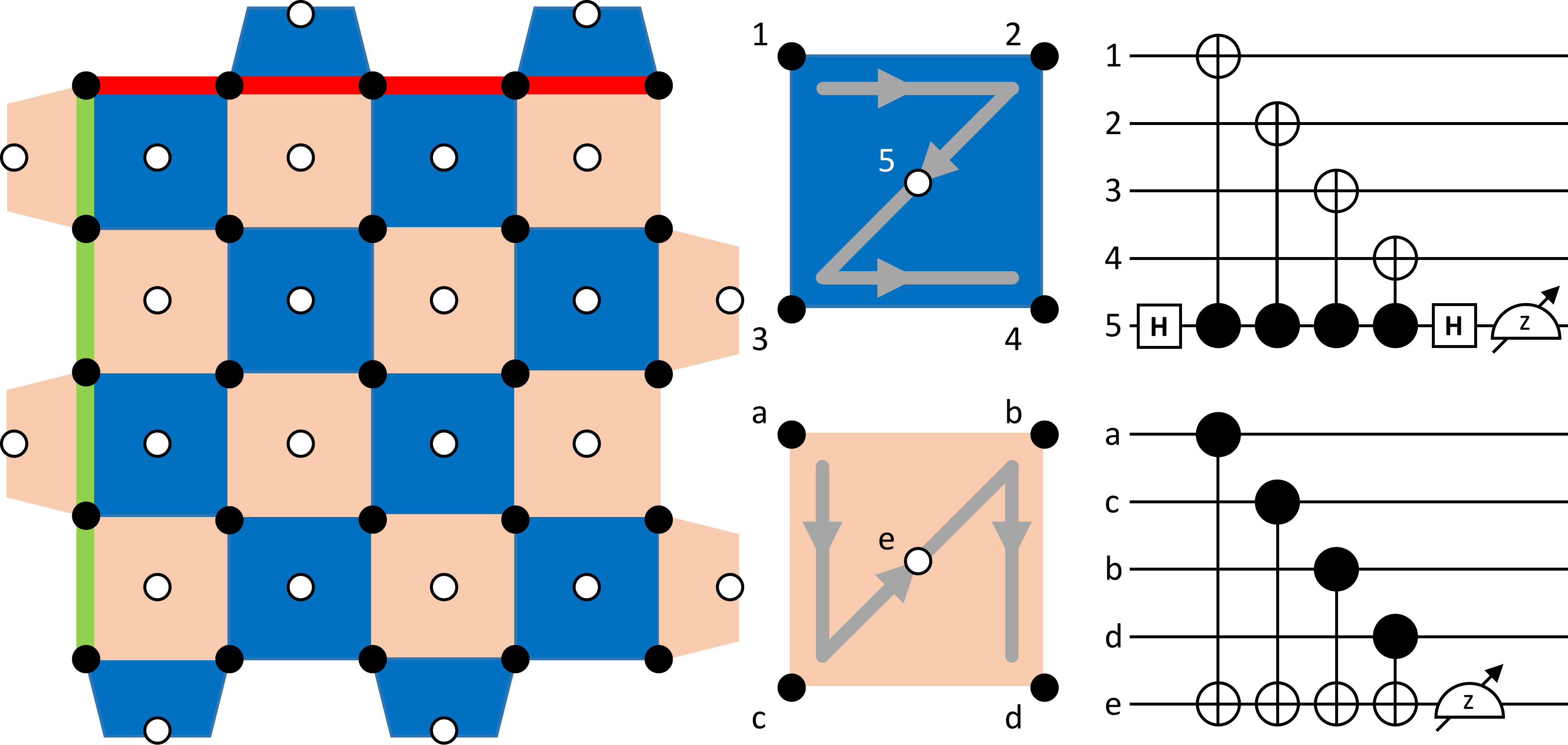}
    \caption{A rotated surface code patch with distance $d=5$. Data (ancilla) qubits are  black (white) circles, on the vertices (faces). The $X$- ($Z$-) stabilizers are  dark (light) faces; Quantum circuits for their measurements are shown. Logical $Z$ ($X$) operator is a red (green) line at the top (left side).}
    \label{fig:surface_code_layout}
\end{figure}

In a rotated surface code patch\cite{Bombin_2007} data qubits are on the vertices of a square grid, while ancilla qubits in the middle of the faces, as shown in Fig.~\ref{fig:surface_code_layout}. 
A logical qubit is encoded in the collective quantum state of the data qubits, more precisely into the so-called logical subspace: the $+1$ eigensubspace of the $X$- and $Z$-stabilizers, called $A_f$ and $B_f$,  
\begin{align}
    A_f &= \prod_{j\in\partial f} X_j; &
    B_f &= \prod_{j\in\partial f} Z_j.
\end{align}
Stabilizers are $X$ and $Z$ parity checks on the data qubits at the corners of the corresponding faces; their measurement can detect local Pauli errors. These measurements are realized via a syndrome extraction circuit which entangles data qubits with ancillas, as in Fig.~\ref{fig:surface_code_layout}. 

Logical operators of the surface code, also shown in Fig.~\ref{fig:surface_code_layout}, can be defined as
\begin{align}
    Z_L &= \prod_{j \in \text{TOP}}Z_j;& 
    X_L &= \prod_{j\in \text{LEFT}}X_j.
\end{align}
These commute with all the stabilizers and act as Pauli operators in the logical subspace. Note that the above definition is not unique: equivalently valid logical operators can be obtained by \change{multiplied}{multiplying} the $Z_L$ and $X_L$ defined above with any stabilizers, since this does not change how they act in the logical subspace. The code distance $d$ is defined as the minimum weight of logical operators, which is just the linear size of the rotated surface code patch. For Fig.~\ref{fig:surface_code_layout}, the distance is $d=5$.

\subsection{Characterizing the noisy logical identity channel}

The noisy logical identity channel describes what happens to an idling logical qubit, if errors affect the physical qubits of the error correcting code. The error model will be detailed in the next Section. 
In the case of incoherent physical errors the channel reads,
\begin{equation}
    \varepsilon(\rho) = (1-p_L)\rho + p^L_xX_L\rho X_L + p^L_y Y_L\rho Y_L + p^L_z Z_L\rho Z_L,
\end{equation}
with the parameters $p^L_{x,y,z}$ describing the probabilities of Pauli X, Y, Z errors. The total \emph{logical failure rate} is $p_L$,
\begin{equation} \label{eq:log_rate}
    p_L = p^L_x + p^L_y + p^L_z.
\end{equation}
Characterization of the noisy logical identity channel is the (numerical) computation of the parameters $p^L_{x,y,z}$, which depend on both the details of the quantum error correcting code and the errors affecting the physical qubits. 

\begin{figure}[!h]
    \centering
    \includegraphics[width = .45\textwidth]{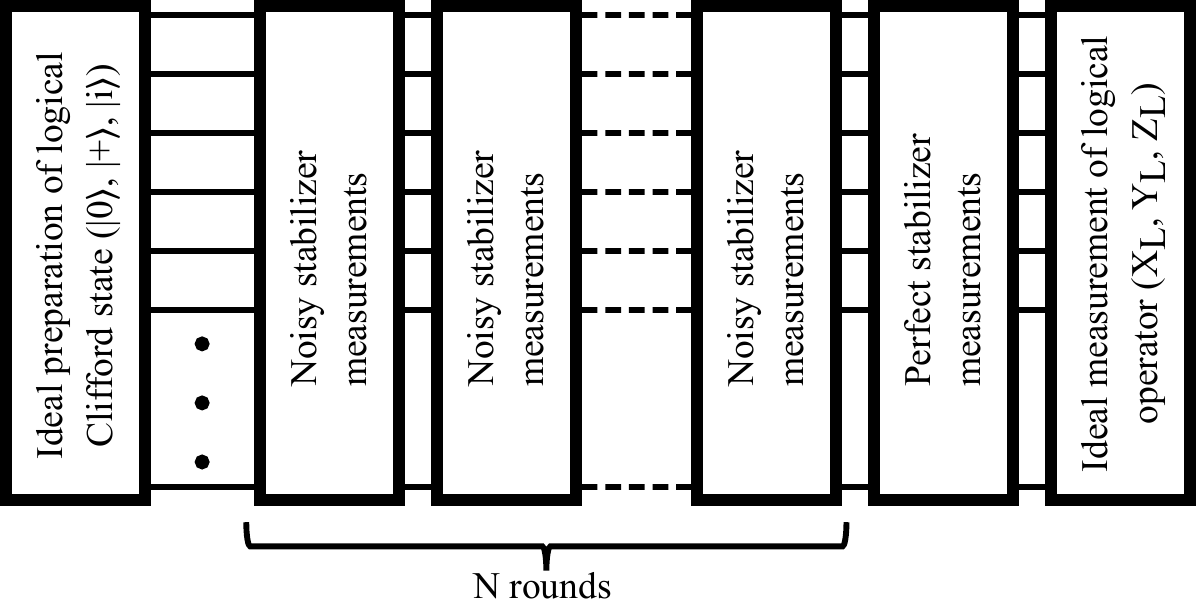}
    \caption{The sketch of quantum circuits realizing ideal $X$, $Y$ and $Z$ memory experiments with a logical qubit encoded into a surface code patch.}
    \label{fig:id_channel_circuit}
\end{figure}

We fully characterize the noisy logical identity channel of a logical qubit encoded in a single surface code patch under circuit-level Pauli noise.
We perform efficient Clifford simulations with the Python package STIM \cite{Gidney_2021_stim} 
of 
ideal memory experiments. The numerical data is available at \cite{marton_2024_13320275}. 
Our approach, as shown in Fig.~\ref{fig:id_channel_circuit}, is as follows. 
We first ideally (with a circuit without noise) initialize the patch in one of the logical Clifford states, $\ket{0}_L,\ket{+}_L$ or $\ket{i}_L$, then measure the stabilizers through noisy syndrome extraction circuits, $N$ times. 
At the end we perform a perfect stabilizer measurement round to ensure the quantum state is left in the logical subspace up to a correction operator (see next paragraph). 
Finally, we measure all of the data qubits in the appropriate basis (in accordance with the initialized state) and extract the eigenvalue of the appropriate logical Pauli operator, $Z_L,X_L$ or $Y_L$\change{}{, which is sufficient to determine the logical failure rate (see Appendix~\ref{apx:logfail})}. \change{}{By performing the noisy stabilizer measurement rounds with the quantum circuits shown in Fig.~\ref{fig:surface_code_layout}, the fault distance of the ideal memory experiments is uniformly $d$}

For the decoding, i.e., to decide a correction operator based on the stabilizer measurement outcomes, we used the minimum weight perfect matching approach \cite{dennis2002topological,fowler2014minimum} via PyMatching \cite{higgott2022pymatching,higgott2023sparse}.
\change{}{We note that other decoders may offer better performance than minimum weight perfect matching \cite{Barber_2025,Bausch:2023xam,Higgot2023,shutty2024efficientnearoptimaldecodingsurface}, but we expect qualitatively similar results with them.}
We used STIM to decompose the detector error model into weighted $X$ and $Z$ matching graphs, where the weights display the probabilities in the circuit-level error model.

Note that we do not model in detail the preparation of the initial state -- this is not relevant for our purpose. In principle, the preparation of logical Clifford states and the measurement of logical Pauli operators can be done fault tolerantly \cite{gidney2023inplace}. However, we would like to characterize the logical identity channel itself and not the noisy memory experiments, so we exclude noise in the initialization and logical measurement processes. 
In other words, our goal in this work is to find the optimal strategy for protecting logical quantum information encoded in idling surface code patches while some other patches may undergo logical operations by any of the proposed approaches \cite{Horsman_2012,Brown_2017,Bravyi_2005,Brown_2020,Litinski_2019,Bombin_2023}.

\section{Error model} \label{sec:error}

\begin{figure*}[!t]
    \centering
    \includegraphics[width = \textwidth]{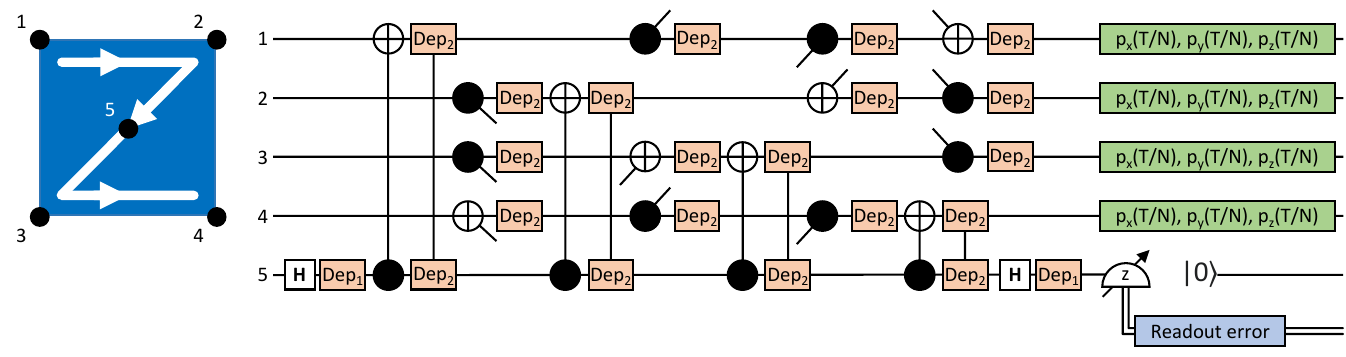}
    \caption{Quantum circuit for the noisy measurement of an $X$-stabilizer. Some CNOT gates are only partly drawn: these are parts of other stabilizer measurements, each free end points towards the ancilla qubit on which the other half of the CNOT acts. Idling errors are green rectangles acting on data qubits during the measurement\change{}{ and the reset} of the ancilla. Readout error is a classical probabilistic bit-flip. Gate errors are orange rectangles.}
    \label{fig:Stab_meas_circuit}
\end{figure*}

We consider three types of errors during quantum error correction rounds, to give a rough description of noise in current quantum devices \cite{google2023suppressing,Bluvstein_2023}:
\emph{idling errors}, describing the dissipative dynamics of data qubits;  \emph{gate errors}, modeling the noisy quantum gates; and \emph{readout errors}, describing imperfect measurements. Among the error types we neglect are leakage errors, correlated errors (these two can be \change{efficiently suppressed}{suppressed to some degree} with various physical-level solutions \cite{Miao2022OvercomingLI,lacroix2023fast,Marques_2023,mcewen2024resistinghighenergyimpactevents}), and crosstalk \cite{Ketterer_2023,Sarovar_2020} (which we plan to investigate in follow-up work). 
We also neglect coherence in the errors by considering only random Pauli error channels. 
This approximation (obtained, e.g., using Pauli twirling) is necessary for the large scale numerical simulations\cite{gottesman1998heisenberg,aaronson_2004}. 
Neglecting coherence in the physical errors is generally believed to be a reasonable approximation, as supported by analytical\cite{beale_2018,Iverson_2020}, numerical \cite{Bravyi_2018,M_rton_2023,pataki2024coherent}, and experimental \cite{google2023suppressing} work. 

\textbf{Idling errors} describe the dissipative dynamics of qubits. They are here modeled based on the amplitude and phase damping channel \cite{Nielsen_Chuang_2010}, characterized by two time scales $T_1$ and $T_{\phi}$: the relaxation and pure dephasing times. The corresponding Lindblad equation in the rotating frame reads:
\begin{align} \label{eq:lindblad}
    \dfrac{\partial \rho}{\partial t} = &\dfrac{1}{T_1}\Big(\ket{0} \bra{1}\rho\ket{1}\bra 0 - \dfrac{1}{2}\{\ket{1}\bra{1},\rho\}\Big) + \nonumber\\
    &\dfrac{1}{T_{\phi}}\Big(\ket{1}\bra{1}\rho \ket{1}\bra{1} - \dfrac{1}{2}\{\ket{1}\bra{1},\rho\}\Big).
\end{align}
After solving Eq.~\ref{eq:lindblad} and applying the Pauli twirl approximation we get
\begin{equation} \label{eq:idling_channel}
    \rho(t) = p_0(t)\rho_0 + p_x(t)X\rho_0 X + p_y(t) Y\rho_0 Y + p_z(t) Z\rho_0 Z,
\end{equation}
with time dependent Pauli error probabilities,
\begin{align} \label{eq:idling_errors}
    p_x(t) &= p_y(t) = \dfrac{1}{4}(1-e^{-t/T_1}) \\
    p_z(t) &= \dfrac{1}{2}(1-e^{-t/T_2}) - \dfrac{1}{4}(1-e^{-t/T_1}) \\
    p_0(t) &= 1-p_x(t)-p_y(t)-p_z(t).
\end{align}
Here $1/T_2 = 1/2T_1 + 1/T_{\phi}$ is the rate of dephasing \cite{Tomita_2014}. 

In the numerical simulations we apply idling errors via Eq.~\ref{eq:idling_channel} on the data qubits, during the time when the ancilla qubits are being measured and reinitialized, see Fig.~\ref{fig:Stab_meas_circuit}. We approximate this time as 
$T/N$ in each cycle, where $T$ is the total time of the of memory experiment, and $N$ is the number of stabilizer measurement rounds. Our basic assumption is that the measurement time is much longer than gate times, which is realistic in current devices
\cite{google2023suppressing,Bluvstein_2023}, thus we neglect the dissipative dynamics of qubits during gate operations.

\textbf{Gate errors} are errors occuring during imperfect quantum gates, whose microscopic model depends very much on the implementation of the gate. 
We thus use a generic approach: 
a single-qubit depolarizing channel after each single qubit quantum gate, and 
a two-qubit depolarizing channel after each two-qubit quantum gate. 
These channels read:
\begin{align}
    \text{Dep}_1(\rho) &= (1-p)\rho + \dfrac{p}{3}\sum_{P\in\{X,Y,Z\}}P\rho P; \\
    \text{Dep}_2(\rho) &= (1-p)\rho + \dfrac{p}{15}\sum_{P\in\{I,X,Y,Z\}^{\otimes 2}/\{II\}}P\rho P,
\end{align}
where $I$ is the single qubit identity operator.

\textbf{Readout errors} model faulty measurements\change{}{ and faulty resets}, which we treat on the phenomenological level: ideal measurements\change{ followed by ideal resets} with possibly erroneously recorded results. 
The measurement outcome $m$ can suffer a classical bit-flip during the readout, therefore, we record the wrong measurement outcome with probability $q$,
\begin{equation}
    P(m=1 \rightarrow m=-1) = P(m=-1 \rightarrow m=1) = q,
\end{equation}
and correspondingly $P(m=1 \rightarrow m=1) = P(m=-1 \rightarrow m=-1) = 1-q$. In Fig.~\ref{fig:Stab_meas_circuit} a segment of the noisy syndrome extraction circuit is shown, displaying all three error mechanisms.

\section{Optimal number of stabilizer measurement rounds} \label{sec:results}

We ran extensive numerical simulations to characterize the logical identity channel for a broad spectrum of parameters: idling error time scales $T_1$, $T_{\phi}$, gate error parameter $p$, readout error probability $q$, and code distance $d$. 
We identified an optimal number of measurement rounds $N^*$, for each set of parameters, where the logical failure rate is minimal. In this Section we 
give a qualitative explanation, and the numerical results, of how $N^*$ depends on different parameters. Finally, we discuss the implications of our results for two promising experimental platforms: superconducting qubits and neutral atoms.

\begin{figure}[!h]
    \centering
    \includegraphics[width=.48\textwidth]{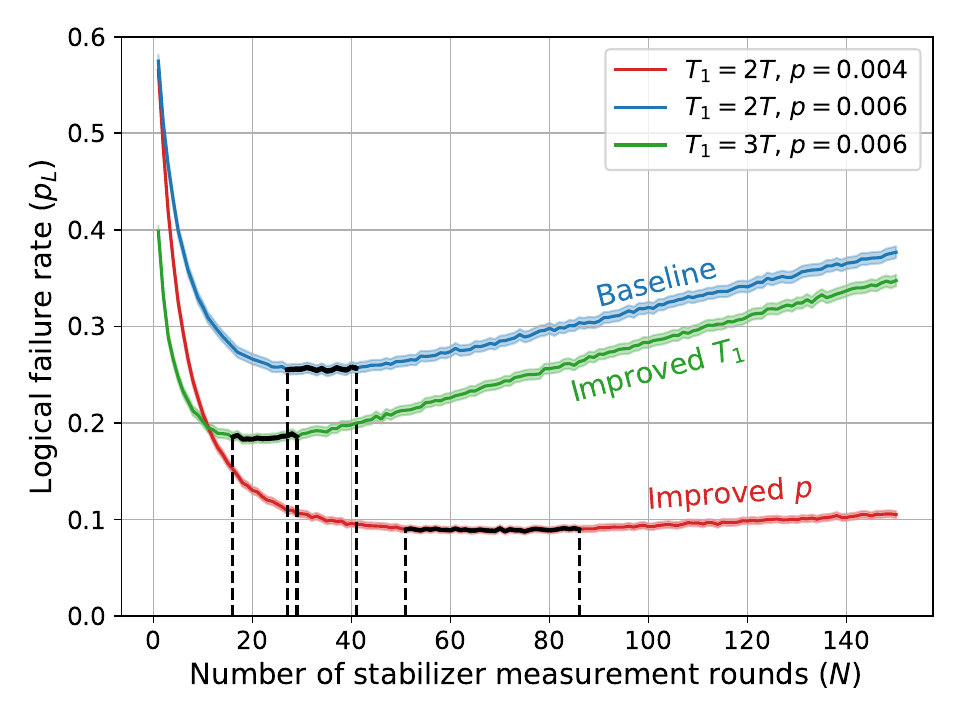}
    \caption{Numerically obtained logical failure rates\change{ $p_L$}{, $p_L = p^L_x + p^L_y + p^L_z$,} for three different sets of  parameters, against the number of measurement rounds, $N$. The pure dephasing time ($T_{\phi}=12T$), the readout error rate ($q=0.02$), and the code distance ($d=9$) were fixed, while the relaxation time ($T_1$) and the gate error rate ($p$) differ for the three curves. The broad minima (minimum to within \change{$\Delta p_L=\sqrt{p_L(1-p_L)}/100$}{$\Delta p_L$}) are highlighted in black. At each stabilizer measurement round value, we ran \change{three}{} ideal \change{}{$X$, $Y$, and $Z$} memory experiments with 100.000 execution rounds for each. \change{}{The error bars are included as pale contours around the curves with width of $6\Delta p_L$.}}
    \label{fig:example_curves}
\end{figure}

We start with a qualitative explanation why one can expect that there is an optimal number of stabilizer measurement rounds, $N^*$, to be performed during a fixed time $T$, to minimize the logical failure rate $p_L$. With few measurement rounds, idling errors dominate the error budget, thus, decreasing $N$ increases the chance that idling errors accumulate to cause an unobserved logical error. 
With many measurement rounds, gate and readout errors dominate the error budget, 
thus, increasing $N$ increases the chance of a logical error. 
Examples of curves showing this behaviour are depicted in Fig.~\ref{fig:example_curves}.

\begin{figure*}[!t]
    \centering
    \includegraphics[width = \textwidth]{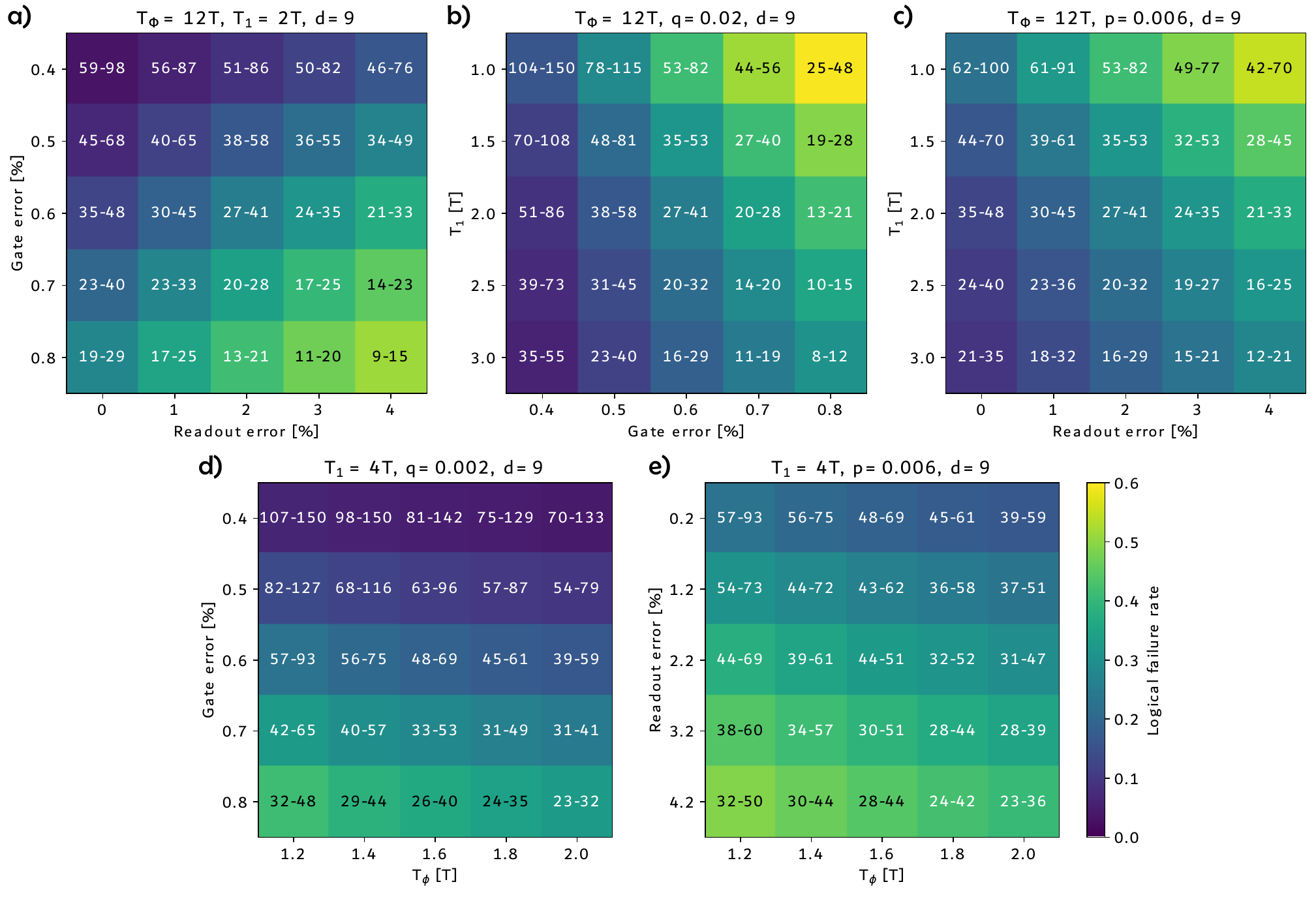}
    \caption{The optimal number of stabilizer measurement rounds and the corresponding logical failure rates for a distance $9$ surface code patch over a wide range of physical error parameters. Numbers indicate the borders of the optimal intervals, while the heatmap indicates the logical failure rates. (a),(b), and (c) correspond to the relaxation-dominated parameter sets, conversely, (d) and (e) to the dephasing-dominated parameter sets.}
    \label{fig:heatmaps}
\end{figure*}

It is quite intuitive to understand qualitatively how the optimal number of stabilizer measurement rounds, $N^*$, depends on the noise parameters. For higher quality qubits (longer $T_1$ and $T_{\phi}$ times) $N^*$ is smaller, because fewer noisy rounds are sufficient to suppress the effect of idling errors. Conversely, for more precise gates and measurements (smaller $p$ and $q$),  $N^*$ is larger, because each stabilizer measurement round introduces less noise, making it beneficial to perform more rounds. Numerical simulations confirm these intuitive expectations as shown in Fig.~\ref{fig:heatmaps}. We also find, as in Fig.~\ref{fig:example_curves}, that the logical failure rate curves are flat around the minima. This results in a large uncertainty in $N^*$, a roughly identical performance in a wide range of $N$.

\begin{figure}[!h]
    \centering
    \includegraphics[width =.48\textwidth]{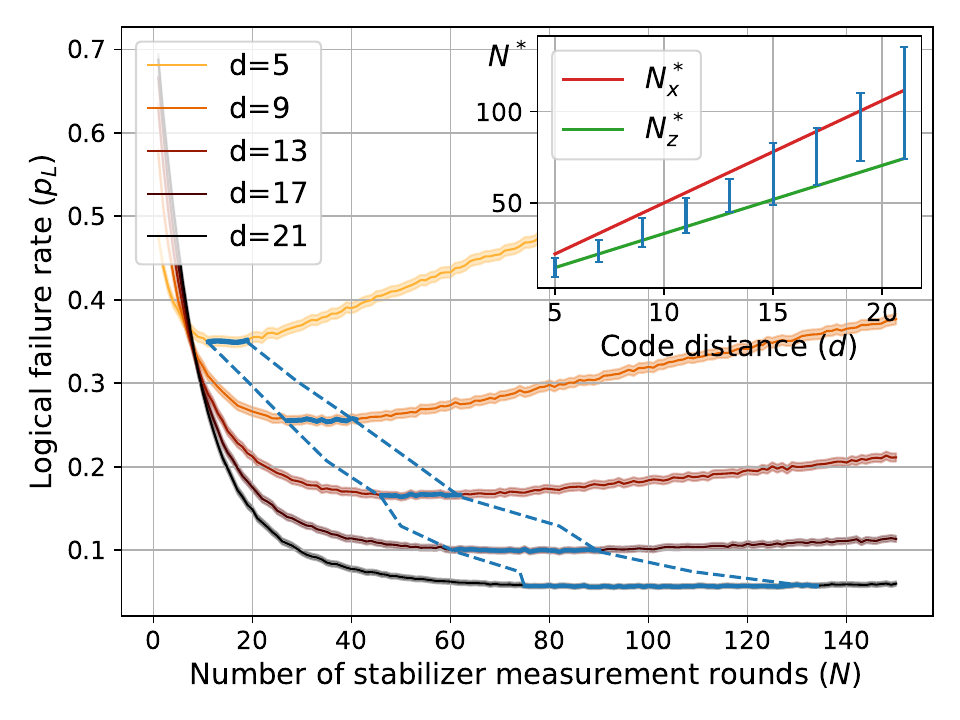}
    \caption{Logical failure rate\change{}{, $p_L = p^L_x + p^L_y + p^L_z$,} as the function of the number of stabilizer measurement rounds for code sizes $d=5$ to $21$. The optimal intervals are colored blue. In the inset, these optimal intervals (blue) can be seen as the function of the code distance. Our estimations in Eq.~\eqref{eq:optimal_numbers} for the optimal number of rounds during memory experiments ($N^*_x$ and $N^*_z$) are shown as red and green curves. The noise parameters were $T_1=2T$, $T_{\phi}=12T$ ,$p=0.006$ and $q=0.02$.\change{}{The error bars are included as pale contours around the curves with width of $6\Delta p_L$.}}
    \label{fig:code_distance_scaling}
\end{figure}

We find that the optimal number of stabilizer measurement rounds also depends on the code distance: $N^*$ increases as the surface code patch is scaled up, see Fig.~\ref{fig:code_distance_scaling}. 
To understand this behaviour we approximate the probability of logical $X$ and $Z$ errors with the contributions from the minimum weight error strings,
\begin{align} \label{eq:error_rate_ansatz}
    p^L_{x/z} &= AN\Big(p_{x/z}(T/N)+p_y(T/N)+kp\Big)^{(d+1)/2}.
\end{align}
Here $A$ is some function of the code distance $d$, but independent of $N$, and $k$ is independent of both $d$ and $N$. The factor $AN$ is the multiplicity of error strings of weight $(d+1)/2$, composed of data qubit errors. These errors originate from idling errors with probabilities $p_x + p_y$ (for logical $X$ errors) or $p_z + p_y$ (for logical $Z$ errors), and from gate errors with probability $kp$. By counting the possible error events (see Appendix~\ref{apx:min_error_strings}) we estimate $k$ as $56/15$. In Eq,~\eqref{eq:error_rate_ansatz} we neglect ancilla qubit errors (originating from readout errors and gate errors affecting ancilla qubits), because these usually do not contribute to minimum weight error strings. A more detailed explanation can be found in Appendix~\ref{apx:min_error_strings}. 

By approximating the idling error rates as 
\begin{align}
    p_{x/z}(T/N)+p_y(T/N) &\approx \dfrac{T}{2T_{1/2}N},
\end{align}
and differentiating $p_x^L$ and $p_z^L$ separately we derive the optimal numbers of stabilizer measurement rounds for $Z$ and $X$ memory experiments,
\begin{align} \label{eq:optimal_numbers}
    N^*_{x/z} = \dfrac{(d-1)T}{4kpT_{1/2}}.
\end{align}
We have plotted $N^*_x$ and $N^*_z$ as the function of $d$ in Fig.~\ref{fig:code_distance_scaling} for the parameter set $T_1=2T$, $T_{\phi}=12T$, $p=0.006$, $q=0.02$, and $k=56/15$.

The optimal number of stabilizer measurement rounds $N^*$ lays between $N^*_x$ and $N^*_z$, and increases with $d$, because both $N^*_x(d)$ and $N^*_z(d)$ are monotonically increasing functions of $d$. Fig.~\ref{fig:code_distance_scaling} confirms this behaviour and validates Eq.~\eqref{eq:optimal_numbers}. A more detailed analysis of the $d$ dependence of the optimal number can be found in Appendix~\ref{apx:code_distance}. In addition,  Eq.~\eqref{eq:optimal_numbers} verifies the intuitive expectations on the $T_1$, $T_{\phi}$, and $p$ dependence of $N^*$. To express the effect of readout errors, ancilla qubit errors have to be taken into consideration in Eq.~\eqref{eq:error_rate_ansatz} as well. 

\subsection{Comparison with Current Experiments}

In the numerical simulations we assumed that the number of measurement rounds can be increased boundlessly. However, in real devices stabilizer measurements take some time to execute, therefore, it may happen that the optimal number cannot be reached in an experiment. 

To compare the theoretically optimal and the experimentally feasible number of stabilizer measurement rounds, we consider two quantum error correction experiments from the last few years, one from Google Quantum AI with superconducting qubits \cite{google2023suppressing} and one from Harvard/QuEra with neutral atoms \cite{Bluvstein_2023}. The error parameters and the execution times of stabilizer measurement cycles of these experiments are summarized in Tab.\ref{tab:experimantal_params}.
\begin{table}[!h]
    \centering
    \begin{tabular}{|c|cc|}
        \hline
         & \shortstack{Superconducting \\ qubits \cite{google2023suppressing}} & \shortstack{Neutral \\ atoms \cite{Bluvstein_2023}} \\
        \hline
        $T_1$ & $20$ $\upmu$s & $4$ s \\
        $T_{\phi}$ & $120$ $\upmu$s & $1.85$ s \\
        $p$ & $0.006$ & $0.006$ \\
        $q$ & $0.02$ & $0.002$ \\
        Cycle time & $\sim900$ ns & $\sim2$ ms \\
        \hline
    \end{tabular}
    \caption{Error parameters and quantum error correction cycle times for superconducting and neutral atom devices. The values are taken from \cite{google2023suppressing,Bluvstein_2023,Bluvstein_2022}. }
    \label{tab:experimantal_params}
\end{table}
In our previous choice of parameters in Fig.~\ref{fig:heatmaps}, 
the relaxation-dominated family of parameter sets ($T_1=1-3T,T_{\phi}=12T,p=0.004-0.008,q=0-0.04$) corresponds to superconducting qubits for experiment time $T=10$ $\upmu$s, 
while the dephasing-dominated family of parameter sets ($T_1=4T,T_{\phi}=1.2T-2T,p=0.004-0.008,q=0.002-0.042$) corresponds to neutral atoms for experiment time $T=1$ s.

While the optimal numbers of stabilizer measurement rounds are of the same order of magnitude for both types of devices (see Fig.~\ref{fig:heatmaps}), the experimentally feasible error correction rounds differ significantly. With the superconductor experiment, under $10$ $\upmu$s at most $11$ stabilizer measurement rounds can be performed, in contrast to that on the neutral atom based quantum device, under $1$ s at most $500$ rounds. Consequently, the optimal strategy differs for these two devices: For $d=9$ (Fig.~\ref{fig:heatmaps}), for superconducting qubits it is beneficial to do the maximum number of measurement rounds with no idling time, however, for neutral atoms, measuring the stabilizers less frequently is suggested\change{ leaving idling time in between for optimal performance.}{. Leaving some idling time in between stabilizer measurements provides optimal performance here.}

Although above we discussed the case of a fix|change{}{ed} idling time $T$, we believe our conclusions regarding the usefulness of idling time between stabilizer measurements are valid more generally. If the errors are rare enough it is reasonable to assume that logical failure rate is proportional to the number of stabilizer measurement rounds with fixed error rates, i.e., 
\begin{align} \label{eq:rare_error_strings}
    p_L = Nf(p_x,p_y,p_z,p,q,d).
\end{align}
Therefore, the logical failure rate over unit time, $p_L/T$, does not explicitly depend on $T$ or $N$, only on the combination $N/T$. As a consequence the optimal frequency of interventions $N^*/T$ minimizes this failure rate, meaning that our findings are valid for arbitrary idling time. This is true using the logical failure rate over unit time as the efficiency metric\change{, assuming the rarity of error strings holds as per Eq.~(16)}{}.  

\section{Conclusion} \label{sec:conclusion}

In this work we numerically characterized the logical noise channel of an idling surface code patch across various code distances, number of stabilizer measurement rounds, and noise parameters. For each set of noise parameters and code distances, we determined an optimal number for stabilizer measurement rounds, $N^*$, that maximizes the performance of the surface code patch during a fixed idling time. Our circuit-level noise model incorporated both time-dependent dynamic errors, amplitude and phase damping of data qubits, and time-independent static gate and readout errors. 

Our results indicate that 
$N^*$ decreases with higher qubit quality, while it increases with higher gate quality and larger code sizes. We also explored the applicability of our findings to superconducting and neutral atom devices, revealing distinct optimal strategies for these two architectures. Specifically, for superconducting qubits, the optimal strategy involves performing the maximum feasible number of stabilizer measurement rounds, whereas for neutral atoms, the optimal number is significantly lower than the maximum feasible.

Including more sophisticated errors, such as leakage \cite{Miao2022OvercomingLI}, crosstalk \cite{Ketterer_2023,Sarovar_2020}, or correlated errors \cite{mcewen2024resistinghighenergyimpactevents} would be an interesting extension of our work, as would considering more parameters to optimize on (e.g., code, code distance, reset after measurements) to get the best performance, as in \cite{chatterjee2024mitsquantumsorcererstone,gehér2024resetresetquestion}.

We believe our results will be valuable for instrumentalists aiming to further optimize the performances of logical qubits encoded into quantum error-correcting codes.

Our data is available at \cite{marton_2024_13320275}.

\section{Acknowledgement}

This research was supported by the Ministry of Culture and Innovation and the National Research, Development and Innovation Office within the Quantum Information National Laboratory of Hungary (Grant No. 2022-2.1.1-NL-2022-00004), and by the Horizon Europe research and innovation programme of the European Union through the HORIZON-CL4-2022- QUANTUM01-SGA project 101113946 OpenSuperQPlus100 of the EU Flagship on Quantum Technologies. We acknowledge the support of the Wigner Research Centre for Physics’ trainee program.  This work was supported by the HUN-REN Hungarian Research Network through the Supported Research Groups Programme, HUN-REN-BME-BCE Quantum Technology Research Group (TKCS-2024/34).

\bibliography{main}
\bibliographystyle{quantum}

\appendix

\section{Minimum weight error strings} \label{apx:min_error_strings}

We approximated the probabilities of logical $X$ and $Z$ errors with the contributions coming from the minimum weight error strings in Eq.~\eqref{eq:error_rate_ansatz}. Here, we provide a detailed explanation of this approximation, and also motivate our choice of neglecting error strings that contain ancilla qubit errors.

Ordering the CNOT gates properly (see Fig.~\ref{fig:Stab_meas_circuit}) does not reduce the code distance at the circuit level. So, at least $d$ error events are required for an undetected logical error. As a consequence the minimum number of independent error events that can cause a logical error (after correction) is $(d+1)/2$. We call sets of error events that lead to a logical error and consist of $(d+1)/2$ elements, minimum weight error strings.

There are two types of minimum weight error strings: the first one contains error events only on data qubits, while the second one contains one error event on an ancilla qubit and $(d-1)/2$ error events on data qubits.
The minimum weight error strings of the second type only count as minimum weight error strings (they only get corrected "wrongly") if the correction has a smaller weight than the error, which happens if the error events on the ancilla qubits have a smaller probability than on the data qubits. For large enough measurement rounds the error events on the ancilla qubits are more likely, therefore, we neglected the contribution of the minimum weight error strings of the second type. However, a more detailed analysis should contain the contribution of these error strings as well.

The error events on data qubits originate from gate or idling errors,
therefore, the probabilities can be written in a form which we already used in Eq.~\eqref{eq:error_rate_ansatz},
\begin{align}
    &p^{x/z}_{data} = p_{x/z}(T/N) + p_y(T/N) + kp + ...,
\end{align}
here we neglected higher order terms. 

The multiplicity of the minimum weight error strings is denoted as $AN$ in Eq.~\ref{eq:error_rate_ansatz}. Here we considered this multiplicity to be  proportional to $N$, for which we used the rarity of error strings as in Eq.~\eqref{eq:rare_error_strings}.

\section{Counting error events}
\label{Apx:counting_errors}

\begin{figure}[!h]
    \centering
    \includegraphics[width = .48\textwidth]{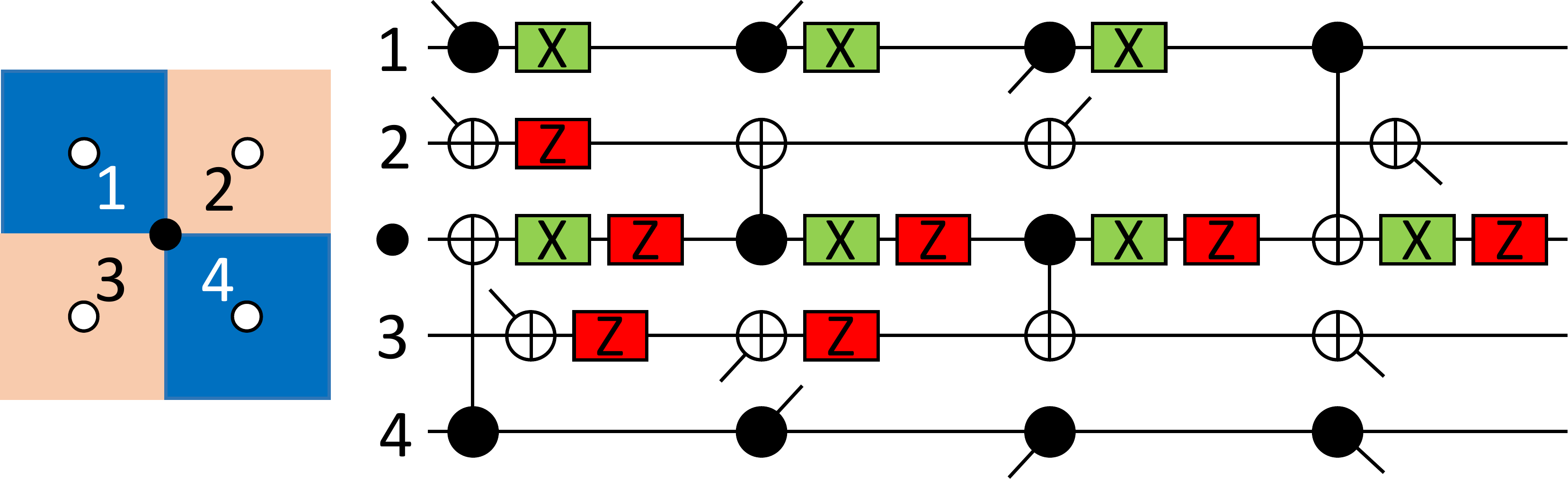}
    \caption{Possible gate error events during a stabilizer measurement round causing an error on a specific data qubit. A segment of the code patch is depicted with a single data qubit (black) and 4 ancilla qubits (white). The relevant part of the syndrome extraction circuit with the possible X (green) and Z (red) error events that cause an error on the data qubit at the end of the given round.}
    \label{fig:error_counting}
\end{figure}

To estimate $k$ (the weight of gate errors in physical error probability) we approximately counted the possible error events that can occur during one stabilizer measurement round. A data qubit participates in 4 CNOT operations which can cause data qubit error events. Also, gate errors on ancilla qubits can propagate to a data qubit from the other 3 CNOT gates. Therefore, the total number of data qubit error events is 7 in both the $X$ and $Z$ cases. Possible error events are shown in Fig.~\ref{fig:error_counting}. The probability of an $X$ or $Z$ data qubit error event, caused by a gate error, is $8p/15$, because $8$ two-qubit Pauli errors consist an $X$ or $Z$ error on a given (first or second) qubit. Therefore,
\begin{align} \label{eq:k_1}
    k = 7 \dfrac{8}{15}=\dfrac{56}{15}.
\end{align}

\section{The $d$ dependence of $N^*$}
\label{apx:code_distance}

\begin{figure}[!h]
    \centering
    \includegraphics[width=0.48\textwidth]{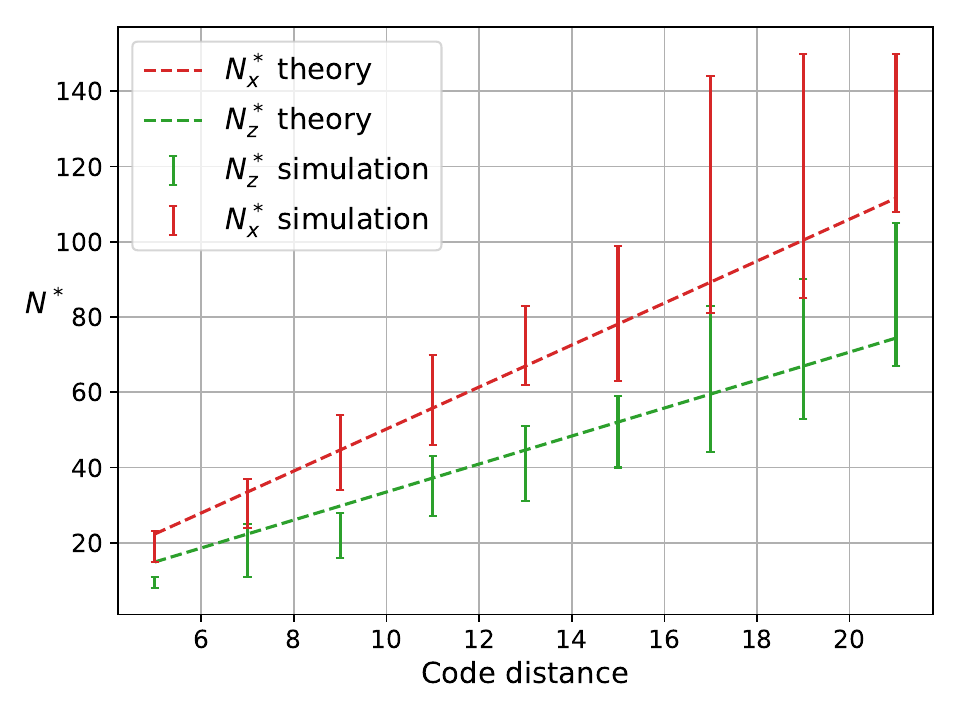}
    \caption{The optimal intervals and the theoretical predictions for the optimal number of measurement rounds separately for $X$ and $Z$ memory experiments as the functions of the code distance.}
    \label{fig:separate_optima_d}
\end{figure}

Analytical expressions can be derived for the optimal number of stabilizer measurement rounds for $X$ and $Z$ memory experiments separately; see Eq.~\eqref{eq:optimal_numbers}. These expressions are derived from the minimum weight error string contributions in the logical failure rate, Eq.~\eqref{eq:error_rate_ansatz}. To see how the theoretically derived expressions of $N^*_x$ and $N^*_z$ match with the numerically determined optimal intervals of X and Z memory experiments, we have plotted these optimal intervals separately for X and Z memory experiments in Fig.~\ref{fig:separate_optima_d}.

\begin{figure}[!h]
    \centering
    \includegraphics[width=0.48\textwidth]{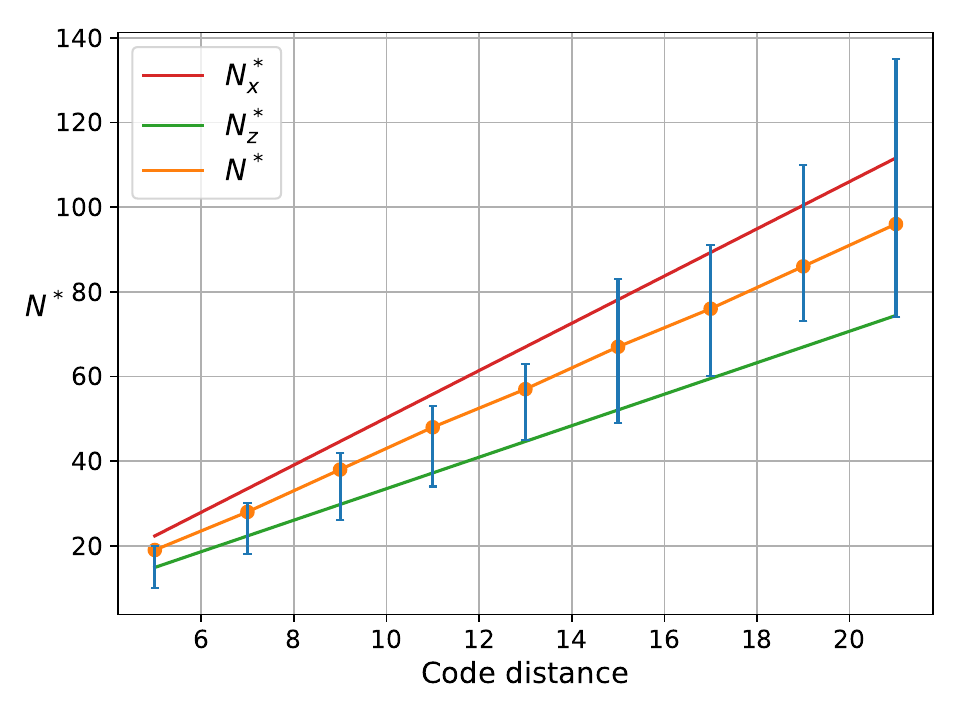}
    \caption{The optimal intervals (blue) and theoretical predictions for the optimal number of measurement rounds as the functions of the code distance. $N^*_x$ and $N^*_z$ are the optimal numbers for the X and Z memory experiments, as described in Eq.~\eqref{eq:optimal_numbers}. $N^*$ is the numerically determined minimum of $p_L = p^L_x+p^L_z + ...$ composed of minimum weight error string contributions as in Eq.~\eqref{eq:error_rate_ansatz}.} 
    \label{fig:true_optima_d}
\end{figure}

Considering the minimum weight contributions in $p^L_x$ and $p^L_z$, as in Eq.~\eqref{eq:error_rate_ansatz}, not only offers analytical expressions for $N^*_x$ and $N^*_z$, but also gives a theoretical prediction for $N^*$, the minimum of $p_L = p^L_x + p^L_z + ...$. We numerically determined the minima of the logical failure rate (contains only minimum weight error string contributions) for the relevant set of parameters and plotted it in Fig.~\ref{fig:true_optima_d}, together with $N^*_x$, $N^*_z$ and the optimal intervals. In Fig.~\ref{fig:true_optima_d} the line of $N^*$ lays between $N^*_x$ and $N^*_z$ and matches with the numerically determined optimal intervals, as mentioned in the main text.

\section{Determining the logical error rate} \label{apx:logfail}

\change{}{To fully characterize the noisy logical identity channel, we simulated ideal memory experiments as discussed in Sec.~\ref{sec:surface_code}. In the absence of noise, the measurement outcome of the appropriate logical Pauli operator at the end of the experiment should be $+1$. However, errors that cause logical failures flip the sign of this final logical measurement. The probabilities of obtaining a $-1$ measurement outcome for the logical Pauli operators can be expressed in terms of the parameters of the noisy logical identity channel:}
\begin{align}
    P(\langle X_L \rangle = -1) &= p^L_y+p^L_z; \\ \nonumber
    P(\langle Y_L \rangle = -1) &= p^L_x+p^L_z; \\ \nonumber
    P(\langle Z_L \rangle = -1) &= p^L_x+p^L_y.
\end{align}
\change{}{Therefore, the logical failure rate, as defined in Eq.~\eqref{eq:log_rate}, can be calculated as:}
\begin{align}
    p_L = \dfrac{P(\langle X_L \rangle = -1) + P(\langle Y_L \rangle = -1) + P(\langle Z_L \rangle = -1)}{2}.
\end{align}
\change{}{We determined the uncertainty in the logical failure rate,}
\begin{align}
    \Delta p_L = \sum_{\mathcal{P}\in X_L, Y_L,Z_L}\dfrac{\Delta P(\langle \mathcal{P} \rangle = -1)}{2} \\
    = \sum_{\mathcal{P}\in X_L, Y_L,Z_L}\sqrt{\dfrac{P(\langle \mathcal{P} \rangle = -1)P(\langle \mathcal{P} \rangle = 1)}{4N}},
\end{align}
\change{}{assuming that the number of failures in each memory experiment follows a binomial distribution. The total number of execution rounds is $N$ for each memory experiment.}

\end{document}